\documentclass[intlimits,twoside,a4paper]{article}

\usepackage{amsmath,amssymb}
\usepackage{graphicx}

\usepackage[T2A]{fontenc}
\usepackage[cp1251]{inputenc}

\usepackage[eqsecnum]{cmpj2}

\issue{2014}{17}{1}{13601}
\doinumber{10.5488/CMP.17.13601}

\title
{Analytical calculation of the Stokes drag of the spherical
particle in a nematic liquid crystal}

\author{M.V. Kozachok\refaddr{label1, label2},
        B.I. Lev\refaddr{label1}}

\addresses{
\addr{label1} Bogolyubov Institute for Theoretical Physics of the National Academy of Sciences of Ukraine, \\14-b
Metrolohichna St., 03680 Kyiv, Ukraine \addr{label2} Open
International University of Human Development ``Ukraine'', 23
Lvivska St., Kyiv, Ukraine}

\date{Received February 8, 2013, in final form July 11, 2013}

\begin{document}
\maketitle
\begin{abstract}

As an approach to the motion of particles in an anisotropic
liquid, we analytically study the Stokes drag of spherical
particles in a nematic liquid crystal. The Stokes drag of
spherical particles for a general anisotropic case is derived in
terms of multipoles. In the case of weak anchoring, we use the
well-known distribution of the elastic director field around the
spherical particle. In the case of strong anchoring, the multipole
expansion may be also used by modifying the size of a particle
to the size of the deformation coating. For the case of zero
anchoring (uniform director field) we found that the viscosities
along the director $\eta_{\parallel}$ and perpendicular direction
$\eta_{\perp}$ are almost the same, which is quite reasonable because in this
case the liquid behaves as isotropic. In the case of non-zero
anchoring, the general ratio $\eta_{\parallel}/\eta_{\perp}$ is
about 2 which is satisfied by experimental observations.


\end{abstract}

\section{Introduction}

Colloidal particles in liquid crystals (LC) have attracted a great
research interest during the recent years. Anisotropic properties of
the host fluid-liquid crystal give rise to a new class of
colloidal anisotropic interactions that never occur in isotropic
hosts. Liquid crystal colloidal systems have shown much recent interest
as the models for diverse phenomena in condense matter physics.
Particles suspended in a fluid are under the effect of the
hits from the surrounding particles and perform Brownian motion.
They perform random walk whose diffusion constant obeys the famous
Stokes-Einstein relation. A simple Langevine approach predicts
that the velocity autocorrelation function of random walkers
decays exponentially \cite{russ}. The drag force can be derived
from the Navier-Stokes equations with an additional assumption on
the character of the random force. The Navier-Stokes equations,
which describe the hydrodynamic behavior of fluids, assume that
molecules are point particles or smooth spheres and, as
a consequence, do not exert a torque on one another. These equations
originate from the conservation of mass, linear momentum and
energy during the collision processes. If the particles in a fluid
are of non-spherical shape, they can induce rotation to each
other during the collisions and the energy can be transferred from
the translational motion to the rotational motion. During these
collisions, the  total angular momentum of colliding particles
should be conserved. The requirement that the angular momentum should be
conserved together with the Navier-Stokes equations leads to a
complete hydrodynamic description of the fluid. Such a complete
hydrodynamic description was applied to the fluid composed of
finite-sized spherical particles with internal rotational degrees
of freedom and it is shown that the friction force becomes memory
dependent even for this simple liquid \cite{rei}.

In anisotropic liquids, the rod-like organic molecules align, on
average, along a common direction indicated by a unit vector
$\vec{n}$ called director. For this case, to find the drag force
we need to solve the dynamic equations of a nematic liquid crystal
LC, i.e., the Ericksen-Leslie equations. In these equations, the
independent variables~--- the director and the fluid velocity~--- are
coupled and this fact causes the complexity of these equations.
Thus, only a few examples with analytical solution exist, e.g., the
flow between two parallel plates which defines the different
Miesowicz viscosities \cite{curr}, the Couette flow
\cite{atk,curri}, the Poiseuille flow \cite{atkin} which was
first measured by Cladis et al. \cite{white}, or back flow
\cite{pier}. It is expected that the knowledge of the more
or less general solutions of these equations will shed light upon
some effects. The solutions of the Ericksen-Leslie equations
are also of technological interest since they are indispensable for
determining the switching times of liquid-crystal displays.

Every particle immersed in a liquid crystal produces a
deformation director field around the particle if the LC
molecules are specifically anchored to the closed surface. In the
case of a weak anchoring, the area of deformation of the
director field around the particle is small and every deformation
of the director field can be presented as a small deformation of
the ground state, which represents the orientation of all the
molecules in one direction.

\begin{figure}[ht]
\centerline{
\includegraphics[width=8.6cm]{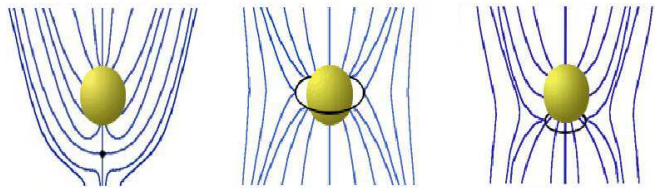}
}
\caption{(Color online) The distortion of the molecules around the spherical
particle in the case of the strong anchoring. We can see that the
change of the distortion of the director near the particle is very
strong. The form of the distortion of the director field in the
case of the strong anchoring was theoretically obtained in article
\cite{kuksenok}.} \label{fig1}
\end{figure}

In the case of a strong anchoring, we have a distortion director
field around the immersed particle, which can be called a dipole or
quadrupole configuration \cite{lubensky} (figure~\ref{fig1}).
 This configuration directly depends on the strength of coupling
with the surface and on the size of particles. In this case, it is
necessary to describe the possible configurations and to note that
in the long-range distance we have a configuration which shows the
same behavior as in the case of week anchoring represented
by a multiple expansion. There exist two approaches to
describe the distribution of the director field at a short and
long distance from the immersed particle. The first theoretical
approach was developed in \cite{lubensky} combining the ansatz
functions for the director and the use of the multiple expansion in the
far field area. The authors investigated spherical particles
with hyperbolic hedgehog and found dipole and quadruple
elastic interactions between such particles. Another approach
\cite{lev} made it possible to find approximate solutions in
terms of the geometrical shape of particles for the case of
small anchoring strength and has provided the way to connect the
type of the interaction potential with the local symmetry of the
director field around the particles \cite{lev1}. The concept of
coating has been introduced that contains all the topological defects
located inside and carries the symmetry of the director, and enables us to
qualitatively determine the type of the interaction potential. However,
the coating is not quantitatively exactly defined. The configuration
of director distribution plays a crucial role when the particle
moves through a liquid crystal.

The hydrodynamic solution for the flow of a nematic liquid crystal
around a particle at rest, which is equivalent to the problem of a
moving particle, still requires its full result. The experiment with the
inverted nematic emulsion \cite{poul,cabuil} and investigations
by Ruhwandl and Terentjev \cite{ruhw} urged Stark and Ventzki
\cite{stark,fukuda,stark2} to perform Stokes drag calculations for
a particle in a nematic environment, especially for the
particle-defect dipole. They concentrated on the low Eriksen
numbers, where the director field is not affected by the velocity
field. The authors presented streamline patterns, interpreted them,
calculated Stokes drags for motions parallel and perpendicular to
the overall symmetry axis, and compared the results to the
Saturn-ring configuration and a uniform director-field.  Heuer et
al. presented analytical and numerical solutions for both the
velocity field and the Stokes drag where the director field was
kept uniform \cite{kneppe,heuer}. They were the first to investigate
a cylinder of infinite length \cite {schneider}. Diogo \cite{diogo} put the velocity field to be the same as the one for
an isotropic fluid and calculated the drag force for simple
director configurations. He investigated the case where the
viscous forces largely exceed the elastic forces from director
distortions, i.e., Ericksen numbers much larger than one, as it
was explained in the \cite {stark}. Roman and Terentjev, have
focused on the opposite case. They obtained an analytical solution for
the flow velocity in a spatially uniform director field by an
expansion in the anisotropy of the viscosities \cite {roman}.
Chono and Tsuji performed a numerical solution of the
Ericksen-Leslie equations around a cylinder determining both the
velocity and director field \cite{chono}. They found that the
director field strongly depends on the Eriksen number, but for
homeotropic anchoring their director fields did not exhibit any
topological defects required by the boundary conditions signaling
about some shortcomings in the exploration. Billeter and Pelcovits
used molecular-dynamic simulations to determine the Stokes drag of
very small particles \cite {billeter}. They observed that the
Saturn ring is strongly deformed due to the motion of the
particles. Ruhwandl and Terentjev have investigated a nonuniform
but fixed director configuration, and numerically calculated
the velocity field and Stokes drag of a cylinder \cite {ruhwa} or
spherical particle \cite {ruhw}. The particle was surrounded by
the Saturn-ring configuration, and the cylinder was accompanied by
two disclination lines.  It is known when a particle is surrounded
by a disclination ring, the Stokes drag strongly depends on the
presence of line defects. There are a few studies that
determine both experimentally \cite{cladis} and theoretically
\cite{imura,genes,ryskin} the drag force of a moving disclination.

We cannot fully describe all the effects associated
with the possible configurations of the director field around the
immersed particle, but we attempt to find a general motive
of the change dissipation energy of the moving particle in a
liquid crystal. First of all, we focus on increasing the
effective mass of the immersed colloidal particle and analytically
calculate the Stokes drag for colloidal particles in a nematic
liquid crystal.

\section{Theory and details of calculations}

The essence of this paper is the calculation of the Stokes drag of
a spherical particle in a nematic liquid crystal when the angle
between the director and particle velocity is arbitrary. In other
words, we have calculated the Stokes drag of the spherical particle
in a nematic liquid crystal for the fully anisotropic case. The
drag force is caused by the interaction of the particles of the
fluid and a foreign body immersed in it. As we mentioned in
the introduction, every particle immersed in a liquid crystal is
dressed in a deformation coating with the region of
deformation of the director field at the distance of the
correlation length. The efficacy of the coating was investigated in
\cite{lev2}. To describe this phenomena we can also use the
results on the inertial characteristic and viscosity, which
present a different approach to the motion of the immersed
particles. The inertial characteristic is the effective mass which
is an analogue of the hydrodynamical mass in the usual
hydrodynamics. Every moving particle immersed in a liquid crystal
has two principal different characteristics. One is an inertial
characteristic as an effective mass and another characteristic
determines the dissipative part. When the particle moves, the
region of deformation~--- i.e., coating, moves too. This causes an
increase of its inertia mass. Under these conditions, the
effective mass becomes the anisotropic value and can be expressed
via formulas \cite{lev2}:
\begin{eqnarray}
m_{\mathrm{eff}}^{\perp} &=&m+I\int \rd\vec{r}\left\{\left(\frac{\partial
\vec{n}}{\partial x}\right)^{2}+\left(\frac{\partial \vec{n}}{\partial y}\right)^{2}\right\}, \\
m_{\mathrm{eff}}^{\parallel} &=&m+I\int \rd\vec{r}\left(\frac{\partial
\vec{n}}{\partial z}\right)^{2},
\end{eqnarray}
where $I$ is density of the moment inertia of the liquid crystal.
To determine the inertial characteristic, we can use the
distribution of the elastic director around the particles. As was
shown in \cite{kuksenok}, in the case of a weak anchoring, when only
small deviations of the director for homeotropic boundary
conditions on the surface of a particle are expected, the problem
can be linearized, and to describe the director field one can use
the two principal angles of a spherical coordinate system
$n_z=\cos\beta(\vec{r}),n_x=\sin\beta(\vec{r})\cos\phi$, and
$n_y=\sin\beta(\vec{r})\sin\phi$, where $\phi$ is the azimuthal
angle, thus respecting an obvious azimuthal symmetry of the
problem. At a small anchoring $\beta\ll1$, the director rotation
angle takes the form
\begin{equation}
\beta=\frac{WR}{4K}\left(\frac{R}{r}\right)^3\sin2\theta.
\end{equation}

If we substitute the known director field distribution for weak
anchoring, we get the value of the inertial effective mass:
\begin{equation}
m_{\mathrm{eff}}=m+\frac{4I}{3}\left(\frac{W}{4K}\right)^{2}R^{3},
\end{equation}
which can be by an order higher than the mass of the immersed particle
\cite{lev2}. It is analogue of the hydrodynamic mass for the moving
particle in an ordinary liquid.

 The friction force for a spherical
region of a radius ${R}/{\varepsilon}$ with a centered particle
within is expressed via formula \cite {stark}.

The same arguments relate to the viscosity coefficient. The
theoretical calculations \cite{fukuda,stark2} revealed that
the viscosity coefficient depends on the configuration of the director
distribution and is much  bigger than in an ordinary viscose
liquid. The essence of this phenomena can be understood from simple considerations. Every particle that moves in the viscose
environment undergoes the action of the additional friction force,
which is described by Stokes formula $f=6\pi{\eta}R$, where $\eta$
is the friction coefficient, which is associated with a diffusion
coefficient of the Brownian particle via the relation
$D={(kT)}/{(6\pi\eta R)}$. The friction force for the spherical
region of radius ${R}/{\varepsilon}$ with a centered particle
within is expressed via formula \cite {stark}
\begin{equation}
f=6\pi\eta
R\frac{1-\frac{3\varepsilon}{2}+\varepsilon^{3}-\frac{\varepsilon^{5}}{2}}{\left(1-\frac{3\varepsilon}{2}+\varepsilon^{3}\right)^{2}}\,.
\end{equation}

From this formula it is easy to see that the friction force increases
if the particle is inside the shell. It can be a solvate shell and
in the case of the liquid crystal this is the region of a  strong
change of a director deformation. If we now take into account the
configuration of the director distribution around the spherical
inclusion, then the diffusion of this inclusion will depend on
the direction of the motion with regard to equilibrium director
distribution. This leads to the anisotropy of the diffusion
coefficient and to a dependence of these coefficients on the conditions
of anchoring on the surface of the inclusion. The results of
numerical calculations of these phenomena can be found in
\cite{fukuda,stark2}.

To determine the character and the value of the Stokes drag of
the spherical particle in a nematic liquid crystal, we can use
different distributions of elastic director field around it.
The stress tensor $\sigma_{ik}$
is used to calculate the Stokes drag force \cite{landau}.
From the known stress tensor $\sigma_{ik}$, the drag force can be calculated
by the following formula
\cite{rei}:
\begin{equation}
F_{i}=\int_{}^{}\sigma_{ij}\rd s_
{j} \,. \label{eq:1.1}
\end{equation}
The expression for the stress tensor $\sigma_{ik}$ in a nematic
environment is well known and can be found in the literature
\cite{landau}
\begin{equation}
\sigma_{ik}=-p\delta_{ik}+\sigma_{ik}^{(r)}+\sigma_{ik}'\,.
\label{eq:1.2}
\end{equation}
Here, $p$ is macroscopic pressure, $\sigma_{ik}^{(r)}$ is
``reactive'' part of stress tensor and $\sigma_{ik}'$ is
a dissipative part of stress tensor. The expressions for ``reactive''
and dissipative parts of stress tensor can be found in \cite{landau}
\begin{eqnarray}
\sigma_{ik}^{(r)}&=&-\pi_{kl}\partial_{i}n_l-\frac{\lambda}{2}(n_ih_k+n_kh_i)+\frac{1}{2}(n_ih_k-n_kh_i),
\label{eq:1.3}
\\
\sigma_{ik}' &=& 2\eta_1v_{ik}+(\eta_2-\eta_1)\delta_{ik}v_{ll}\nonumber\\
&&{}+(\eta_4+\eta_1-\eta_2)(\delta_{ik}n_ln_mv_{lm}+n_in_kv_{ll})\nonumber\\
&&{}+(\eta_3-2\eta_1)(n_in_lv_{kl}+n_kn_iv_{il})\nonumber\\
&&{}+(\eta_5+\eta_1+\eta_2-2\eta_3-2\eta_4)n_in_kn_ln_mv_{lm}\,.
\label{eq:1.4}
\end{eqnarray}

To find the stress tensor we need to know the solution of the
Eriksen-Leslie equations that link the director field and the
fluid velocity. The general solution of these equations is a
challenge to a theorist.

Here we suggest an approach for finding the stress tensor. As
the first step we use the director structure around a colloid
particle suspended in a nematic liquid crystal, found in \cite{kuksenok}. We assume here the situation when a spherical
particle moves slowly and the nematic liquid crystal environment
has enough time to relax to the equilibrium state during the
motion of a spherical particle. We consider a smooth hard sphere
which moves through the fluid with the velocity $
{\vec{u}}(t)=u(t)\vec e_{z}$. The fact that it is
smooth means that no torques and no force directed tangent to its
surface can be exerted on it. Under these conditions, only the
component $\sigma_{rr}$ of the stress tensor contributes to the
drag force. Since we use the director field for equilibrium
state, the ``reactive'' part of stress tensor will not contribute to
the drag force, but only a dissipative part. It is obvious that the
drag force $\vec {F}$ has the same direction as the velocity
of a spherical particle $\vec {u}$, and formula
(\ref{eq:1.1}) will reduce to the following:
\begin{equation}
F=\int\sigma'_{rr}\cos{\theta}\rd s.
 \label{eq:1.6}
\end{equation}

Substituting the components of director field from \cite{kuksenok}
and the components for velocity field which are the same as the
one for an isotropic fluid \cite{lifh} in the stress tensor and
keeping terms up to the first order of small parameter $\beta$, we
have obtained the Stokes drag of spherical particle in a nematic
environment at weak anchoring
\begin{eqnarray}
F &=& 4\pi{}R\eta_1u+3\pi{}Ru(\eta_1+\eta_2-\eta_3-\eta_4)\left(-0.27-0.02\frac{WR}{K}\right)\nonumber\\
&&{}+3\pi{}Ru(\eta_5+\eta_1+\eta_2-2\eta_3-2\eta_4)\left(0.11+\frac{WR}{K}\right).
 \label{eq:1.7}
\end{eqnarray}

The presented director field structure contains configurations of
the director field at a small anchoring. We would like to
determine the Stokes drag in the case of strong anchoring. The
task of finding a director distribution around a spherical
particle consists in minimizing the Frank free-energy
functional with boundary conditions provided by it and by the
surface energy. Generally, this class of problems is not solvable
analytically due to its nonlinearity brought in by the
unit-vector constraint ${\vert\vec {n}(\vec r)\vert}^2=1.$
 In
\cite{gennes}, in particular, the director distribution was
obtained in the one-constant approximation in terms of the
multipole expansion. However, the expansion coefficients were not
associated with the physical and geometrical parameters of
macroparticles. In \cite{lev}, the director distribution is
derived for the general case of different elastic Frank constants
and, moreover, the multipole expansion parameters are found in
terms of geometric and physical characteristics of macroparticles.
Thus, both the behavior and the value of the pair interaction
energy are described with no additional restrictions. However, only in
 \cite{lubensky} there was proposed a theoretical approach combining
the ansatz functions for the director field and the use of the multiple
expansion in the far field area that was a satisfactory solution to many problems.
Thus, the use of the director field in
terms of the multipole expansion becomes particularly relevant. In
the framework of this approach we have
\begin{equation}
n_{x}=p_{z}\frac{x}{r^3}+2c\frac{zx}{r^5}\,,\qquad
 n_{y}=p_{z}\frac{y}{r^3}+2c\frac{zy}{r^5}\,,\qquad
n_{z}=1-\frac{1}{2}\left(n_{x}^2+n_{y}^2\right),
 \label{eq:1.8}
\end{equation}
where the vector $\vec {p}$ is the dipole moment of the
droplet-defect configuration and the parameter $\emph{c}$ is the
amplitude of the quadrupole moment tensor $\emph{c}_{ij}$ of the
particle-defect configuration. Assuming that, at small anchoring,
the director deviates from its uniform orientation
$\vec n_{0}\Vert\vec e_{z}$ by only a small amount, we can consider
$n_x$ and $n_y$ as small parameters. Repeating all the steps as in
the first approach, except that now the small parameters are $n_x$
and $n_y$, we obtain the Stokes drag of a spherical particle in a
nematic environment at a weak anchoring in the framework of the present
approach
\begin{eqnarray}
F &=& 4 \pi R \eta_1 u +(\eta_4 +\eta_3 -\eta_2 -\eta_1)\left[\cos^{2} 2\Omega \cdot \frac{0.06 u
\pi{}c}{R^{2}}+\cos 2 \Omega \cdot \left(\frac{0.18 u \pi{}
c}{R^{2}} +0.6 u \pi{} R \right)+0.2 u \pi{} R\right]\nonumber\\
&&{}+(\eta_5 +\eta_1 +\eta_2 -2\eta_3-\eta_4)\left[\left(1+\sin2\Omega\right)\left(-0.08\cos^{4}\Omega+0.44\cos^{2}\Omega-0.12\sin\Omega -0.24\right)
\left(\frac{u\pi{}c}{R^{2}}\right)\right.\nonumber\\
&&{}+\left(\frac{u\pi{}c}{R^{2}}\right)\left(0.17\cos2\Omega+0.6\cos^{6}\Omega-0.12\cos4\Omega
-0.06\cos6\Omega\right) +3u\pi
R\left(0.08\cos^{2}\Omega\cdot\sin^{2}\Omega\right.\nonumber\\
&&{}\left.\left.-0.06\sin^{4}\Omega\right) + 0.06 u\pi R \sin^{2}2\Omega +0.165 u\pi R
\left(3\cos^{4}\Omega-\cos^{2}\Omega\right)%
\vphantom{\left(\frac{u\pi{}c}{R^{2}}\right)}
\right].
 \label{eq:1.9}
\end{eqnarray}
Here, $\Omega$ is the angle between the director $\vec{n}_{0}$
and the velocity of a spherical particle $\vec {u}$.
This is a general expression for the drag force which includes
anisotropy. We can note that it is a very good approximation for
a spherical particle with the coating size $R$. Outside this
region, there are only small deformations and we should take
into account the multiple explanation. If we assume that the director
field is equal to zero we see that expressions (\ref{eq:1.7}) and
(\ref{eq:1.9}) become the same, which confirms the rightness of our
calculations. In practice, only two directions are measured i.e., along  and perpendicular to the
director direction. We rewrite the expression (\ref{eq:1.9}) for
these two directions
\begin{eqnarray}
F_{\parallel} &=&
4\pi{}\eta_{1}u+(\eta_{4}+\eta_{3}-\eta_{2}-\eta_{1})\left(\frac{0.24u\pi{}c}{R^{2}}+0.8u\pi{}R\right)\nonumber\\
&&{}+(\eta_5+\eta_1+\eta_2-2\eta_3-2\eta_4)\left(\frac{0.71u\pi
c}{R^{2}}+0.33u\pi{}R\right),
 \label{eq:1.10}
\\
F_{\perp} &=& 4\pi \eta_{1} u+(\eta_{4}+\eta_{3}-\eta_{2}-\eta_{1})\left(\frac{-0.12u\pi c}{R^{2}}-0.4 u \pi R\right)\nonumber\\
&&{}+(\eta_5 +\eta_1 +\eta_2 -2\eta_3 -2\eta_4)\left(\frac{-0.59 u
\pi c}{R^{2}}-0.18 u \pi R\right). \label{eq:1.11}
\end{eqnarray}
Alternative investigation was recently conducted by
\cite{mondiot}. In \cite{mondiot} the authors have developed a
perturbative approach to the Leslie-Ericksen equations and related
the diffusion coefficients to the Miesovicz viscosity parameters
$\eta_{i}$. We present our results in the same order as in
\cite{mondiot} in table~1. The value of quadrupole moment is used
as in \cite{lubensky}. Two cases are considered, i.e., uniform director
field and planar anchoring.
\begin{table*}
\caption{The viscosities for zero and planar anchoring. The middle
column is obtained from equations~(\ref{eq:1.10}) and (\ref{eq:1.11})
with the Leslie coefficients of 5CB and MBBA \cite{prost,oswald};
the last one is derived from Stark's numerical calculations
\cite{stark1}.} \label{table}\vspace{2ex}
\begin{center}
\begin{tabular}{|p{2.7cm}p{2.7cm}p{2.7cm}p{2.7cm}|}
\hline
\multicolumn{4}{|l|}{\hspace{1pt}Uniform director\hspace{79pt}Present work\hspace{28pt}Numerically exact\,}\\
\hline\hline
 $\rm 5CB$ & $\eta_{\parallel}$ & $\rm 0.27 $ & $\rm 0.38$\\
{} & $\eta_{\perp}$ & $\rm 0.24 $ & $\rm 0.75$\\
$\rm MBBA$ & $\eta_{\parallel}$ & $\rm 0.28$ & $\rm 0.38$\\
{} & $\eta_{\perp}$ & $\rm 0.27$ & $\rm 0.68$\\
\hline
\multicolumn{4}{|l|}{\hspace{1pt}Planar anchoring\hspace{89pt}\hspace{28pt}\,}\\
 $\rm 5CB$ & $\eta_{\parallel}$ & $\rm 0.16 $ & $\rm {-}$\\
{} & $\eta_{\perp}$ & $\rm 0.34 $ & $\rm {-}$\\
$\rm MBBA$ & $\eta_{\parallel}$ & $\rm 0.18$ & $\rm {-}$\\
{} & $\eta_{\perp}$ & $\rm 0.36$ & $\rm {-}$\\
\hline
\end{tabular}
\end{center}
\end{table*}

\section{Conclusion}
For uniform director field, our results differ from the results in
\cite{mondiot,stark1}, but we believe that our results are
reasonable. Our arguments are as follows: in case of a uniform
director field, the field of the director is the same in space. The
difference in the description of the dynamics of the usual liquid
and the liquid crystal is in the expression of the free energy
\cite{landau}. The expression for the free energy for the liquid
crystal contains, in comparison with the expression for the free
energy for an ordinary liquid, an additional term, i.e., deformation free
energy which depends only on the derivatives of the director
regarding the position \cite{landau}. In the case of a uniform
director field, this term becomes zero and the liquid crystal behaves
as an ordinary liquid. Thus, the viscosities along the director
$\eta_{\parallel}$ and perpendicular direction $\eta_{\perp}$
should be the same. Our results confirm this fact in contradiction
to the results of \cite{mondiot,stark1}. The possible
explanation of this discrepancy might be in the fact that the
director evolution depends on the molecular field and the gradient
of the velocities \cite{landau}. A situation is possible when
the molecular field is zero but the gradient of velocities is big
enough (the particle moves quickly) and, consequently, the deviation of
the director is not as small as in our approach. If the anchoring is
not zero, we find that for both kinds of liquid crystals the ratio
$\eta_{\parallel}/\eta_{\perp}$ is about 2 which agrees well with
the general tendency \cite{mondiot}. The approach used by us is
valid for a weak anchoring when $n_x$ and $n_y$ are small
parameters. However,  it may be applied to the case of strong anchoring
as well. As was shown in \cite{lev1,lev2,lev3}, under strong anchoring,
the effective mass of an ion increases due to the formation of a
polarization coating, moving together with the ion. Thus, we can
consider the particle with the coating to be a new single moving particle. The
anchoring for this ``new particle'' is weak \cite{lev1,lev2,lev3}
and the above approach can be applicable too. It should be noted that while extracting the viscosity, the size of the polarization coating should
be taken into account. We can conclude that our approach works
well for the two limiting cases, i.e., weak and strong anchoring, and it does
not include the case of the ``middle'' anchoring. In \cite{mondiot} there
was measured the ratio $D_{\parallel}/D_{\perp}\approx 4$. This
might be the case of the ``middle'' anchoring. We cannot claim a complete theory of  motion of immersed particles in a
nematic liquid crystal, but we suggest the approach which makes
the analytical calculation of the diffusion process of
a particle in this viscose media possible. This process should take into
account the change as an inertial effect and the Stokes drag of
a particle in a liquid crystal which are linked with the deformation
of the elastic director field.


\ukrainianpart

\title{Аналітичний розрахунок сили Стокса для сферичної частинки в нематичному рідкому кристалі}
\author{М.В. Козачок\refaddr{label1, label2}, Б.І. Лев\refaddr{label1}}
\addresses{
\addr{label1} Інститут теоретичної фізики ім. М.М. Боголюбова НАН
України, Україна, Київ, вул. Метрологічна, 14-б
 \addr{label2} Відкритий міжнародний університет розвитку людини ``Україна'',
Україна, Київ, вул. Львівська, 23}
%
%
%

\makeukrtitle

\begin{abstract}
Як одне з наближень до опису явища руху частинки в анізотропній
рідині, ми аналітично рахуємо силу Стокса для сферичних частинок в
нематичному рідкому кристалі. Сила Стокса для сферичної частинки
порахована для загального анізотропічного випадку в термінах
мультипольного розкладу. Для випадку слабкого зчеплення ми
використовуємо добре відомий розподіл поля директора навколо
сферичної час\-тин\-ки. Для випадку сильного зчеплення мультипольний
розклад також можна використовувати, якщо модифікувати розмір
частинки до розміру шуби. При нульовому зчепленні (однорідне поле
директора) ми знайшли, що вязкість вздовж та перпендикулярно до
директора є однаковою. Цей результат є правдоподібним, оскільки в
цьому випадку рідкий кристал веде себе як ізоторопна рідина. Для
випадку ненульового зчеплення загальне співвідношення
$\eta_{\parallel}/\eta_{\perp}$ є близько 2, що задовільняє
експериментальні результати.

\keywords сила Стокса, рідки кристали, дифузія, вязкість

\end{abstract}

\end{document}